\documentclass[aip,jcp,10pt,twocolumn]{revtex4-1}

\usepackage{amsmath}
\usepackage{amssymb}
\usepackage{amscd}
\usepackage[pdftex]{graphicx}
\usepackage{nicefrac}
\usepackage[np,autolanguage]{numprint}
\usepackage{bm}
\usepackage{siunitx}
\usepackage{multirow}
\usepackage{adjustbox}
\usepackage{siunitx}
\usepackage{color}
\usepackage{physics}
\usepackage{subcaption}
\usepackage{chemformula}
\usepackage{subcaption}

\begin{document}

\title{Performance of a minimally empirical local-hybrid density functional for molecular chemistry}

\author{Erin R. Johnson}
\email{erin.johnson@dal.ca}
\affiliation{
Department of Chemistry, Dalhousie University, 6274
Coburg Rd, Halifax, Nova Scotia, B3H 4R2, Canada}
\affiliation{
Yusuf Hamied Department of Chemistry, University of Cambridge, Lensfield Road, Cambridge, CB2 1EW, UK.}

\author{Kyle R. Bryenton}
\affiliation{
Department of Chemistry, Dalhousie University, 6274
Coburg Rd, Halifax, Nova Scotia, B3H 4R2, Canada}

\date{\today}

\begin{abstract}

Delocalisation error has been argued to be the greatest outstanding challenge in density-functional theory (DFT). One of the most promising routes to minimise this error is development of local hybrid functionals, in which the fraction of exact-exchange mixing is position dependent. However, existing local hybrids capable of good thermochemical accuracy are often highly empirical, and tend to have complicated functional forms that involve some combination of range separation, calibration functions, power-series expansions, or even neural networks. In this work, we explore the limits of what can be achieved with a minimally empirical local hybrid functional form that avoids such complexities. The ``LHnz'' functional is proposed, which uses dispersionless exchange and a local mixing fraction dependent on the correlation length and effective exchange--correlation hole normalisations. With only three empirical parameters, LHnz is shown to outperform all existing global hybrid functionals for the GMTKN55 molecular-thermochemistry benchmark with no large outliers.

\end{abstract}

\maketitle

\section{Introduction}

Kohn--Sham density-functional theory (DFT) has emerged as a dominant method for predicting the electronic structure of chemical systems.\cite{haunschild2016evolution} Its success stems from its balance of accuracy and computational efficiency, allowing it to predict chemical properties of interest at a practical computational cost.\cite{borges2021quantum} It gained widespread popularity when the B3LYP hybrid functional, a modification of that originally proposed by Becke,\cite{beck1993density, lee1988development, stephens1994ab, vosko1980accurate} was shown to approach the accuracy of the G2 method at a fraction of the computational cost.\cite{curtiss1997assessment} 

In theory, DFT is a formally exact method. In practice, however, errors are incurred due to the density functional approximations (DFAs) used to model the exchange--correlation functional, whose exact form is not known.\cite{becke2014perspective, maurer2019advances} One may ascend Perdew's Ladder of DFAs,\cite{perdew2001jacob} where higher rungs include more ingredients with which to capture the physics of the system. Doing so can (hopefully) increase accuracy, but at the cost of additional computational requirements. 

The fourth rung of Perdew's Ladder pertains to hybrid functionals, which incorporate some fraction of exact exchange into the DFT energy. Global hybrid functionals are the most common form of hybrid DFA, with the amount of exact-exchange energy mixing (typically between 20\% and 50\%) fixed globally via
\begin{equation}
E_{\text{XC}} = a_{\text{X}} E_{\text{X}}^{\text{HF}} + (1-a_{\text{X}}) E_{\text{X}}^{\text{DFT}} + E_{\text{C}}^{\text{DFT}} \,,
\end{equation} 
where $a_{\text{X}}$ is the hybrid mixing parameter. Range-separated hybrids, such as the $\omega$B97 family,\cite{chai2008systematic, chai2008long, lin2013long, mardirossian2014omegab97x} have been shown to be extremely powerful for modelling molecular systems.\cite{wtmad4, xdmz} These functionals determine the exact-exchange mixing fraction from the inter-electron distance, replacing the Coulomb interaction with 
\begin{equation}
\frac{1}{r_{12}} = \frac{\text{erfc}(\omega r_{12})}{r_{12}} + \frac{\text{erf}(\omega r_{12})}{r_{12}} \,,
\end{equation}
where $\textrm{erf}$ and $\textrm{erfc}$ are the error function and its complement, and $\omega$ is the range-separation (RS) parameter that determines switching between short- and long-range exchange treatments. However, the range-separation parameter has been shown to be highly system dependent, limiting the transferability of these methods,\cite{bredas,isborn,whittleton} and there is growing evidence that the performance limit of range-separated hybrid functionals has been reached.\cite{liang2026reaching}

Demand for higher-accuracy DFT is constantly growing, and the use of double-hybrids from the fifth rung remains prohibitively costly for large systems. One promising way forward is the development of local hybrid functionals, where the exact-exchange mixing fraction is position dependent: 
\begin{equation}
E_\text{XC} = \int 
g(\mathbf{r}) 
\varepsilon_\text{X}^\text{HF}(\mathbf{r}) \, d\mathbf{r} + 
\int \left[ 1-g(\mathbf{r}) \right] \varepsilon_\text{X}^\text{DFT} (\mathbf{r}) \, d\mathbf{r} + E_\text{C}^\text{DFT} \,,
\end{equation}
where $g(\mathbf{r})$ is the local mixing function. This allows greater exact-exchange fractions in regions of high electron delocalization, potentially improving the accuracy of such systems.\cite{wires} While not explicitly written in this form, real-space correlation models\cite{becke2003real, jaramillo2003local} are also local hybrid functionals due to their dependence on the exact-exchange density, $\varepsilon_\text{X}^\text{HF}$.
While local hybrid functionals show incredible promise for pushing past the aforementioned performance limit, their widespread use has been hindered by the difficulty of implementing computation of the exchange-energy density, which is not available in most production codes.

To assess the accuracy of these methods, the standard benchmark database for molecular chemistry is GMTKN55,\cite{goerigk2017look} which contains 55 benchmarks spanning small and large molecules, reaction barriers, and non-covalent interactions (though more expansive benchmarks exist\cite{liang2025gold}). Due to the wide range of energy scales among the component benchmarks, the overall error across the dataset is typically reported using a weighted mean absolute deviation, such as the new WTMAD-4.\cite{wtmad4} The very best global and range-separated hybrids have weighted errors on the order of \mbox{3--5} kcal/mol.\cite{wtmad4, xdmz} While early local hybrids were hard-pressed to approach this level of accuracy,\cite{kaupp-earlylh} more recently developed local hybrids, such as $\omega$LH25tdE-D4,\cite{kaupp-beyondzero} B22,\cite{b22} and B22plus,\cite{b22plus} have achieved and even surpassed it. However, most local hybrids are highly empirical and/or have complex functional forms. For example, B22 and B22plus have 9 and 20 fit parameters, respectively, while $\omega$LH25tdE-D4 has 21. Several local hybrids also have their mixing functions defined by a neural network (NN).\cite{kaupp2025data,kaupp2026local} 

Our recent work has sought to assess the limits of accuracy achievable with minimally empirical GGA and global hybrid functionals, involving only \mbox{1--2} fit parameters that typically appear only in the London dispersion energy term.\cite{xdmz} This DFA design philosophy is consistent with Occam's razor and is based on capturing essential physics to maximise transferability and avoid overfitting to specific training data. 
Indeed, the recent data-driven direction of DFT development has been called into question, with concerns that researchers are straying from the path towards a near-exact density functional in favour of increasing parametrisation and empirical fitting.\cite{medvedev2017density} In this vein, the present work proposes a simple local hybrid functional with only 3 parameters. The construction explicitly avoids calibration functions and B97-like power-series expansions\cite{becke1997density} that are common in many local hybrids. Assessing its performance on GMTKN55 shows it to already be on par with the best global hybrids and close to that of local hybrids with several times more parameters. 

\section{Theory}

The total density-functional energy of an chemical system can be written as
\begin{equation}
E_\text{DFT} = E_\text{class} + E_\text{XC}\,,
\end{equation}
where $E_\text{class}$ is the classical energy, consisting of the sum of the electrons' kinetic energies, the electron--nuclear attraction energy, the classical electron--electron repulsion energy, and the nuclear repulsion energy. $E_\text{XC}$ is the exchange--correlation energy and accounts for all non-classical contributions. The general form of the exchange--correlation functionals considered in this work is
\begin{eqnarray}\nonumber
E_\text{XC} & = & E_\text{X}^\text{B86b} + \sum_\sigma \int g_\sigma(\mathbf{r}) \left[ \varepsilon_{\text{X},\sigma}^\text{exact}(\mathbf{r}) - \varepsilon_{\text{X},\sigma}^\text{B86b}(\mathbf{r}) \right] d\mathbf{r} \\
 & & + E_\text{C}^\text{PBE} + E_\text{disp}^\text{XDM(Z)}\,,
\label{eq:exc} 
\end{eqnarray}
where $E_\text{X}^\text{B86b}$ is the B86b exchange energy, $E_\text{C}^\text{PBE}$ is the PBE correlation energy, $E_\text{disp}^\text{XDM(Z)}$ is the XDM(Z) London dispersion energy,\cite{xdmz} $\varepsilon_\text{X}^\text{exact}$ and $\varepsilon_\text{X}^\text{B86b}$ are the exact (Hartree-Fock) and B86b exchange-energy densities, respectively, $\sigma$ denotes electron spin, and $g(\mathbf{r})$ is a mixing function. In the B86bPBE0 global hybrid, $g_\sigma(\mathbf{r}) = 0.25$. 

\begin{figure}
\includegraphics[height=\columnwidth,angle=90]{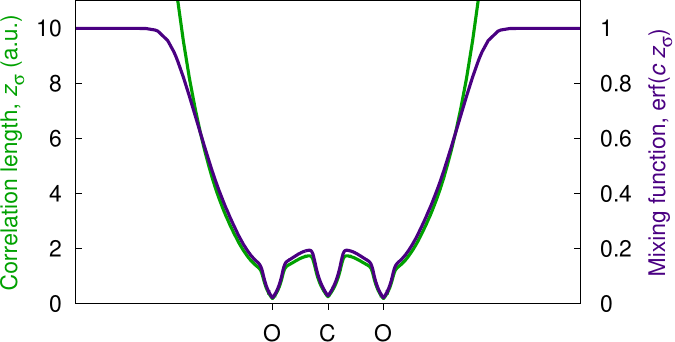}
\caption{Values of the correlation length (in bohr atomic units) and local-hybrid mixing function (with $c=0.10$) along the bond axis of the CO$_2$ molecule.}\label{f:co2}
\end{figure}

Following our earlier work,\cite{johnson2014} we propose a new ``LHz'' local hybrid functional of the form given by Eqn.~\ref{eq:exc}, in which $g_\sigma(\mathbf{r})$ is a function of only the $\sigma$-spin correlation length, $z_\sigma$:
\begin{equation}\label{eq:erf}
g_\sigma(\mathbf{r}) = \erf\{c\, z_\sigma(\mathbf{r})\} \,,
\end{equation} 
where the value of $c$ will be determined later. The correlation length itself is defined as
\begin{equation}
z_\sigma(\mathbf{r}) = \left| \frac{\rho_\sigma(\mathbf{r})}{\varepsilon_{\text{X},\sigma}^\text{B86b}(\mathbf{r})} \right| \,,
\end{equation} 
where $\rho_\sigma$ is the $\sigma$-spin electron density. The correlation length adopts small values near atomic nuclei and large values in the exponentially decaying tails of the electron density, as shown for the example of the CO$_2$ molecule in Figure~\ref{f:co2}. An advantage of this choice of mixing function is that it is smoothly varying and avoids the rapid oscillations near atomic nuclei seen with ratios of kinetic-energy densities.\cite{kaupp-nogauge}

Note that there is not a unique definition of the exact exchange-energy density since one can add any term, $G(\mathbf{r})$, that integrates to zero and still recover the exact exchange energy:
\begin{equation}
E_\text{X}^{\text{exact}} = \sum_\sigma \int \varepsilon_{\text{X},\sigma}^\text{exact} (\mathbf{r})  d\mathbf{r} = \sum_\sigma \int \left[ \varepsilon_{\text{X},\sigma}^\text{exact} (\mathbf{r}) + G(\mathbf{r}) \right] d\mathbf{r}\,.
\end{equation}
Such changes of gauge are achieved by addition of calibration functions to the conventional exact exchange-energy density used in some local hybrids, particularly those with mixing functions that depend on the kinetic-energy density.\cite{lhleftright,lhcore} 
However, it was previously shown\cite{johnson2014} that using the correlation length in the mixing function did not require a change of gauge to obtain good accuracy for thermochemical benchmarks. More recently, Kaupp and co-workers\cite{kaupp-nogauge} proposed a local hybrid with a mixing function based on a ratio of exact and density-functional exchange-energy densities that also did not require a gauge term. As a result, the conventional gauge for the exact exchange-energy density is used throughout this work, and this is in line with the philosophy of designing simple, minimally empirical functionals.

\begin{figure}
\includegraphics[width=\columnwidth]{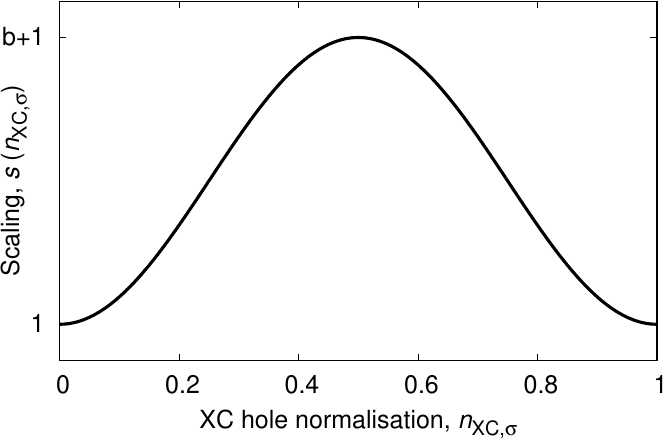}
\caption{Illustration of the scaling function of Eqn.~\ref{eq:scaling}.}\label{f:sin2}
\end{figure}

Returning to Eqn.~\ref{eq:erf}, 
introducing further position dependence can augment the local exact-exchange mixing in situations where delocalisation error is important. In the proposed ``LHnz'' local hybrid, the mixing function argument is enhanced by a further multiplicative term that depends on the exchange--correlation hole normalisation, $n_{\text{XC},\sigma}$:
\begin{equation}
g_\sigma(\mathbf{r}) = \erf\left\{c \, s[n_{\text{XC},\sigma}(\textbf{r})] \, z_\sigma(\mathbf{r})\right\} \,.
\end{equation}
The particular form used for this scaling function is
\begin{equation}\label{eq:scaling}
s(n_{\text{XC},\sigma}) =  b \sin^2(\pi \, n_{\text{XC},\sigma})+1 \,,
\end{equation}
which is designed to equal unity when the effective hole normalisation is an integer (1 or 0) and to increase smoothly for fractional hole normalisations, reaching a maximum value of $(b+1)$ at $n_{\text{XC},\sigma}=0.5$, as shown in Figure~\ref{f:sin2}.
Given this choice, the mixing function used in LHnz becomes
\begin{equation} \label{eq:LHnz_g}
g_\sigma(\mathbf{r}) = \erf\left\{c \left[ b \sin^2(\pi \, n_{\text{XC},\sigma})+1\right] z_\sigma(\mathbf{r})\right\} \,.
\end{equation}

Finally, the effective XC hole normalisation is defined using the 2003 model of Becke,\cite{b03}
\begin{equation}
n_{\text{XC},\sigma} = n_{\text{X}\sigma} + f n_{\text{X}\sigma^\prime}  \,,
\end{equation}
where the exchange-hole normalisations, $n_{\text{X}\sigma}$ and $n_{\text{X}\sigma^\prime}$ are determined using the inverse-Becke--Roussel procedure\cite{b03} and 
\begin{equation}
f = \min \left( \frac{1-n_{\text{X},\alpha}}{n_{\text{X},\beta}}, \frac{1-n_{\text{X},\beta}}{n_{\text{X},\alpha}} , 1\right) \,.
\end{equation}
The purpose of the min function is to maximally deepen the XC hole subject to the constraints that neither the $\alpha$ nor $\beta$ hole normalisation can exceed unity, nor can more than the full $n_{\mathrm{X}\sigma^\prime}$ be added. For the vast majority of closed-shell systems, $n_{\text{XC},\sigma}=1$ so the sine term vanishes and the mixing function becomes identical to that of Eqn.~\ref{eq:erf}. However, for many open-shell systems, particularly in cases with significant delocalisation error, $n_{\text{XC},\sigma}<1$ and the sine term will increase the extent of exact-exchange mixing. 

To illustrate the effect of the mixing term, consider the simple model systems listed in Table~\ref{t:atoms}. Each system corresponds to the dissociation limit of the specified diatomic molecule, which can be viewed as a single atom, with a fractional charge in some cases.\cite{cohen-deloc} 
For the neutral hydrogen and helium atoms, the exchange--correlation holes have unit normalisation, such that the scaling function is simply one. However, for the fractionally charged hydrogen and helium atoms that correspond to the dissociation limits of H$_2^+$ and He$_2^+$, the $\alpha$- or $\beta$-spin XC hole normalisation, respectively, is 0.5. This leads to increased scaling of the error function argument, resulting in enhanced exact-exchange mixing and a reduction in delocalisation error.\cite{wires} 

Reduction of delocalisation error in global and range-separated hybrids is often described as a zero-sum game.\cite{zerosum,kauppaccount} Improved accuracy for fractional-charge systems may be achieved by increased exact-exchange mixing, but at the expense of greater errors for fractional-spin systems such as the spin-unpolarised dissociation limit of H$_2$. However, in the the case of the LHnz functional, the scaling function remains unity for a neutral, spin-averaged hydrogen atom. Thus, LHnz gives the same H$_2$ dissociation limit as LHz, which does not involve the scaling function, avoiding the zero-sum game. While a strong-correlation term\cite{b13,kp16,kauppsc} would still need to be added to LHnz to recover the exact spin-unpolarised H$_2$ dissociation limit and provide accurate treatment of systems with both fractional-spin and fractional-charge errors, this is beyond the scope of the present work.

\begin{table}
\caption{Hole normalisations and scaling-function values for model systems consisting of simple diatomics stretched to their dissociation limits. For the H atom, both the spin-polarised and spin-averaged cases are considered.}\label{t:atoms}
\centering
{\footnotesize
\setlength{\tabcolsep}{2pt}
\begin{tabular}{llcccccc}\hline
Molecule & Atom & $n_{\text{X},\alpha}$ & $n_{\text{X},\beta}$ 
& $n_{\text{XC},\alpha}$ & $n_{\text{XC},\beta}$ &  
$s(n_{\text{XC},\alpha}$) &  $s(n_{\text{XC},\beta}$) \\ \hline
\multicolumn{8}{c}{Integer Charge} \\
H$_2$    & H           & 1   & 0   & 1   & 0   & 1     & 1 \\
H$_2$    & H           & 0.5 & 0.5 & 1   & 1   & 1     & 1 \\
He$_2$   & He          & 1   & 1   & 1   & 1   & 1     & 1 \\ \hline
\multicolumn{8}{c}{Fractional Charge} \\
H$_2^+$  & H$^{+0.5}$  & 0.5 & 0   & 0.5 & 0   & $b+1$ & 1 \\
He$_2^+$ & He$^{+0.5}$ & 1   & 0.5 & 1   & 0.5 & 1     & $b+1$ \\ \hline
\end{tabular}}
\end{table}

\section{Data Sets}

Two data sets were used in this work. The first is \textbf{KB49}, which consists of basis-set extrapolated CCSD(T) binding energies of 49 molecular dimers.\cite{kannemann2010van,xdmhybrid} Dimer geometries are available from the \texttt{refdata} GitHub repository.\cite{otero2015refdata} This set was used to fit the $z_\text{damp}$ parameter that appears in the XDM(Z) damping function\cite{xdmz} by minimising the root-mean-square percent error (RMSPE) in the computed binding energies.

The second data set is \textbf{GMTKN55}, which comprises 55 individual benchmarks spanning the thermochemistry of small and large molecules, reaction barriers, and both intramolecular and intermolecular non-covalent interactions.\cite{gmtkn55zoo}  Geometries (in FHI-aims format) may be obtained from the \texttt{gmtkn55-fhiaims} GitHub repository.\cite{johnson2024gmtkn55} However, the postG code used to evaluate the local-hybrid exchange energy is not yet compatible with effective core potentials (ECPs).\cite{b22,b22plus} As ECPs are required for the HEAVY28 benchmark to include scalar relativistic effects, this benchmark is omitted for all calculations performed in this work and we will instead consider performance for a reduced ``GMTKN54'' set. Similarly, the HAL59 and HEAVYSB11 benchmarks are truncated to eliminate any reference data involving elements heavier than krypton, reducing the sets to 29 halogen bonding energies and 7 heavy-element bond energies, respectively, following previous literature.\cite{b22,b22plus}

Due to the wide range of energy scales across the component benchmarks within GMTKN55, the overall error is reported as a weighted mean absolute deviation (\mbox{WTMAD}). While several definitions have been proposed, this work uses WTMAD-2, which weights by the magnitudes of the reference energies,\cite{gmtkn55zoo} and WTMAD-4, which weights by the magnitudes of the errors for a set of 10 reference, minimally empirical functionals (10-DFA) that serve as a consensus average.\cite{wtmad4,xdmz} 
The WTMAD-2 is defined as
\begin{equation}
\text{WTMAD-2} =  \sum_{i=1}^{N_\mathrm{bench}}  \frac{N_{\mathrm{sys},i}}{N_{\mathrm{total}}} \, 
\frac{|\overline{\Delta E}|_\mathrm{mean}}{|\overline{\Delta E}|_i}  \, \mathrm{MAD}_i \,,
\end{equation}
where the mean energy for each benchmark is
\begin{equation}
|\overline{\Delta E}|_\mathrm{mean} = \frac{1}{N_\mathrm{bench}} \sum_{i=1}^{N_\mathrm{bench}} |\overline{\Delta E}|_i  \,.
\end{equation}
However, this scheme results in some benchmarks contributing to the weighted error $>100 \times$ more than others (e.g.\ IL16 at 0.06\% vs.\ BH76 at 9.89\%), with the top 3 benchmarks contributing as much to the WTMAD-2 as the bottom 36.
We previously proposed WTMAD-4,\cite{wtmad4} which weights benchmarks based on expected errors from a set of 10 minimally empirical hybrid
functionals:
\begin{equation}\label{eq:WTMAD4}
\text{WTMAD-4} = \frac{1}{N_\mathrm{bench}} \sum_{i=1}^{N_\mathrm{bench}}  
w_i \, \mathrm{MAD}_i \,,
\end{equation}
where
\begin{equation}\label{eq:wi}
 w_i = \frac{100}{N_\mathrm{bench}} \left( \frac{3.5}{\overline{\mathrm{MAD}}_i^\text{10-DFA}} \right) \,.
\end{equation}
The 10 reference functionals chosen, all paired with D3 dispersion are: PBE0,\cite{adamo1999toward} B3LYP,\cite{beck1993density,lee1988development,stephens1994ab,vosko1980accurate} BHLYP,\cite{perdew1986density} B3P86,\cite{beck1993density,perdew1986density} B3PW91,\cite{beck1993density,perdew1991electronic} PW1PW91,\cite{perdew1991electronic,adamo1999toward} MPW1PW91,\cite{perdew1991electronic,adamo1998exchange}
HSE06,\cite{krukau2006influence} HISS,\cite{henderson2007importance} and LC-$\omega$hPBE.\cite{vydrov2006assessment} With WTMAD-4, all benchmarks are weighed typically within a factor of 3, and the top 3 benchmarks contribute as
much as the bottom 6.\cite{wtmad4,xdmz}

When calculating the weighted error of our reduced ``GMTKN54'' set, there is some subjectiveness as to how to best modify the WTMAD-2 and WTMAD-4 definitions. To preserve the WTMAD-2 weights to as close as literature definitions as possible, $|\overline{\Delta E}|_\text{mean}=56.84$ will remain a fixed quantity. For the same reason, $|\overline{\Delta E}|_i$ and $N_{\text{sys},i}$ will remain unmodified despite the omission of several reactions with heavy elements from the HAL59 and HEAVYSB11 benchmarks. However, because the HEAVY28 benchmark was excluded in its entirety, the total number of benchmarks is changed to $N_\text{bench}=54$ and total number of systems is reduced to $N_\text{total}=1477$. For WTMAD-4, the weights in Eqn.~\ref{eq:wi} will be preserved exactly, and $N_\text{bench}=54$ when used in Eqn.~\ref{eq:WTMAD4}.

An outlier analysis, as previously shown in Ref~\citenum{xdmz}, is also performed. Here, we analyse the maximum errors and track the number of benchmarks with large errors relative to the mean MAD of the 10 functionals used as the ``consensus average'' in the definition of the WTMAD-4 weights.
$N_{r>h}$ is the number of ratio outliers satisfying $\nicefrac{\text{MAD}_i}{\,\overline{\text{MAD}}_{i}^{\text{10-DFA}}}>h$, while $N_{d>h}$ is the number of difference outliers satisfying $\text{MAD}_i- \overline{\text{MAD}}_{i}^{\text{10-DFA}}> h$ in units of kcal/mol. Ideally $N_{r>1} = N_{d>0}$ would equal zero, meaning the functional has equivalent or better performance than the consensus average in all cases. Barring that, we seek to avoid any extreme outliers, taken to be cases where the error more than doubles relative to the consensus average (i.e.\ $r>2$), or exceeds it by more than 2 kcal/mol (i.e.\ $d>2$). 

\section{Computational Methods} \label{ss:compmethods}

Single-point energy calculations were performed on all systems using the PBE0 functional\cite{perdew1996generalized,adamo1999toward} as implemented in the Gaussian 16 program\cite{g16} to obtain the self-consistent electron densities in the form of  wavefunction (.wfn) files. Following Ref.~\citenum{gmtkn55zoo}, the def2-QZVP basis set\cite{def2basis} was used in all cases, except for benchmarks containing anions (viz.\ AHB21, BH76, BH76RC, G21EA, IL16, WATER27), where it was augmented with diffuse s functions for H, and diffuse s and p functions for all other elements, taken from the aug-cc-pVQZ basis.\cite{augbasis,augbasisheavy} The `Extralink=L608' keyword was used to output the classical ($E_\text{class}$) and PBE correlation energy ($E_\text{C}^\text{PBE}$) components to the total DFT energies. 
The XDM(Z) dispersion energy\cite{xdmz} was computed from the PBE0 electron density using the postg code.\cite{postg}
All exchange energy terms were also computed from the PBE0 electron density using a modified version of Becke's postG code described in Ref.~\citenum{b22}. 

\section{Results and Discussion}

As defined,\cite{xdmz} B86bPBE0-XDM(Z) involves only one empirical parameter, $z_\text{damp}$, which appears in the dispersion damping function. The number of parameters then increases to two for LHz and three for LHnz. The value of $z_\text{damp}$ was optimised for each functional by minimising the root-mean-square percent error for the binding energies of the KB49 set of molecular dimers. The resulting values and error statistics are shown in Table~\ref{t:damping}. Note that LHz and LHnz have identical $z_\text{damp}$ as LHnz reduces to LHz for closed-shell systems, which form the entirety of the KB49 set. The value of the $c$ parameter in LHz and LHnz (Eqs.~\ref{eq:erf} and \ref{eq:LHnz_g}) was set to $c=0.10$ as this was found to minimise the WTMAD-4 for the subset of 41 benchmarks that contain only closed-shell systems. Finally, the value of the $b$ parameter in LHnz (Eqs.~\ref{eq:scaling} and \ref{eq:LHnz_g}) was set to $b=4.6$ to minimise the overall WTMAD-4 for the GMTKN54 set. In keeping with Becke's ``fine tuning'' philosophy,\cite{becke2014perspective} the values determined for $c$ and $b$ are physically intuitive. We expect $c \ll \nicefrac{1}{2}$ because large exact-exchange fractions are not appropriate for describing interstitial regions between bonding atoms. As shown in Fig.~\ref{f:co2}, $z_\sigma \approx 2$ in these regions for the example of CO$_2$. Similarly, the exact exchange fraction should increase from approximately 25\% to near 100\% in regions of fractional charge, suggesting that $b\sim 4$ is appropriate.

\begin{table}[ht!]
\caption{Optimum XDM(Z) damping parameters (in Hartree) for specified functionals implemented post-SCF, using PBE0/def2-QZVP densities. Also shown are the corresponding mean absolute error (MAE, in kcal/mol) and mean absolute percent error (MAPE, in \%) for the KB49 set of intermolecular binding energies.}
\label{t:damping}
\centering
\setlength{\tabcolsep}{6pt}
\begin{tabular}{l|ccc} \hline
Functional & $z_\text{damp}$ & MAE & MAPE \\ \hline
B86bPBE0 & 105695 & 0.33 & 8.7 \\
LH(n)z   & 117237 & 0.21 & 6.5 \\ 
\hline
\end{tabular}
\end{table}

\begin{table}[ht!]
\caption{Comparison of MADs (in kcal/mol) for the individual GMTKN54 benchmarks using the selected functionals with XDM(Z). Also shown for comparison are the $\overline{\text{MAD}}_{i}^{\text{10-DFA}}$ values used in the definition of the WTMAD-4, which may be used to quantify outliers.}\label{tab:shaded}
\centering
\footnotesize
\begin{tabular}{lrrrr}\hline
Benchmark & 10-DFA & B86bPBE0 & LHz & LHnz \\ \hline
AL2X6     &  1.94 &  1.28 &  1.13 &  1.13 \\
ALK8      &  4.88 &  3.03 &  1.94 &  1.94 \\
ALKBDE10  &  6.03 &  5.83 &  5.38 &  5.60 \\
BH76RC    &  2.49 &  1.94 &  2.17 &  2.36 \\
DC13      &  8.16 &  7.39 &  6.55 &  6.12 \\
DIPCS10   &  3.86 &  2.92 &  3.78 &  3.78 \\
FH51      &  2.66 &  2.37 &  2.08 &  2.08 \\
G21EA     &  2.97 &  2.67 &  3.58 &  3.51 \\
G21IP     &  3.92 &  3.71 &  4.56 &  4.65 \\
G2RC      &  5.67 &  6.06 &  4.95 &  4.95 \\
HEAVYSB11 &  2.33 &  1.32 &  2.32 &  1.89 \\
NBPRC     &  2.74 &  2.33 &  1.76 &  1.76 \\
PA26      &  3.21 &  3.23 &  1.65 &  1.65 \\
RC21      &  4.40 &  4.76 &  3.85 &  3.63 \\
SIE4x4    & 13.60 & 14.21 & 13.88 &  6.31 \\
TAUT15    &  1.10 &  1.15 &  1.24 &  1.24 \\
W4-11     &  5.92 &  3.64 &  4.38 &  4.41 \\
YBDE18    &  2.50 &  1.46 &  2.21 &  1.87 \\ \hline
BSR36     &  3.14 &  1.26 &  0.88 &  0.88 \\
C60ISO    &  5.34 &  2.33 &  2.86 &  2.86 \\
CDIE20    &  1.11 &  1.17 &  0.98 &  0.98 \\
DARC      &  4.41 &  2.20 &  2.04 &  2.04 \\
ISO34     &  1.42 &  1.24 &  1.09 &  1.09 \\
ISOL24    &  3.19 &  2.05 &  1.95 &  1.95 \\
MB16-43   & 19.30 & 13.16 & 13.31 & 13.15 \\
PArel     &  1.16 &  1.08 &  0.96 &  0.96 \\
RSE43     &  1.21 &  1.43 &  1.11 &  1.19 \\ \hline
BH76      &  4.17 &  3.69 &  3.06 &  2.60 \\
BHDIV10   &  3.83 &  4.05 &  2.94 &  2.94 \\
BHPERI    &  2.78 &  2.05 &  1.67 &  1.67 \\
BHROT27   &  0.58 &  0.59 &  0.53 &  0.53 \\
INV24     &  1.44 &  1.23 &  1.26 &  1.26 \\
PX13      &  5.40 &  5.44 &  5.61 &  5.61 \\
WCPT18    &  3.59 &  3.21 &  2.98 &  2.98 \\ \hline
ADIM6     &  0.17 &  0.18 &  0.35 &  0.35 \\
AHB21     &  1.02 &  1.16 &  0.50 &  0.50 \\
CARBHB12  &  1.16 &  1.28 &  1.00 &  1.00 \\
CHB6      &  1.54 &  1.43 &  1.11 &  1.11 \\
HAL59     &  0.66 &  0.58 &  0.60 &  0.60 \\
IL16      &  0.55 &  0.30 &  0.98 &  0.98 \\
PNICO23   &  0.81 &  0.54 &  0.45 &  0.45 \\
RG18      &  0.21 &  0.12 &  0.18 &  0.18 \\
S22       &  0.47 &  0.40 &  0.21 &  0.21 \\
S66       &  0.35 &  0.32 &  0.18 &  0.18 \\
WATER27   &  4.44 &  4.90 &  2.63 &  2.63 \\ \hline
ACONF     &  0.10 &  0.07 &  0.07 &  0.07 \\
Amino20x4 &  0.28 &  0.25 &  0.26 &  0.26 \\
BUT14DIOL &  0.21 &  0.19 &  0.13 &  0.13 \\
ICONF     &  0.32 &  0.38 &  0.36 &  0.36 \\
IDISP     &  2.54 &  1.66 &  0.69 &  0.69 \\
MCONF     &  0.30 &  0.29 &  0.55 &  0.55 \\
PCONF21   &  0.74 &  0.79 &  0.53 &  0.53 \\
SCONF     &  0.33 &  0.25 &  0.33 &  0.33 \\
UPU23     &  0.63 &  0.55 &  0.63 &  0.63 \\ \hline
WTMAD-2   &  6.33 &  5.55 &  5.10 &  4.90 \\
WTMAD-4   &  6.36 &  5.44 &  5.30 &  5.19 \\ \hline
\end{tabular}
\end{table}

Table~\ref{tab:shaded} shows the computed MAEs for each of the 54 benchmarks considered, divided according to basic properties and reaction energies for small systems (18), reaction energies for large systems and isomerisations (9), reaction barrier heights (7),  intermolecular noncovalent interactions (11), and intramolecular noncovalent interactions (9). Recall that the HEAVY28  benchmark is omitted from the set of intermolecular noncovalent interactions. The table provides results from the mean of the 10-DFA set used to define the WTMAD-4 weights, which serve as representative errors for each benchmark. Also shown are the MAEs obtained with the B86bPBE0-XDM(Z) global hybrid and the two new local hybrids. 

The results for B86bPBE0-XDM(Z) in Table~\ref{tab:shaded} differ somewhat from those reported previously in Ref.~\citenum{xdmz}. This is due to the use of PBE0/def2-QZVP densities as opposed to self consistent B86bPBE0/tight densities (augmented by diffuse functions where appropriate). The largest differences occur for DARC, WATER27, and BSR36, all of which have MADs that are lower by in excess of 1~kcal/mol using the Gaussian basis sets. This implies that the tight numerical-atomic-orbital basis\cite{abbott2025roadmap} may not be sufficiently well converged for some benchmarks, although it may also be a result of improved error cancellation between the basis set and parametrisation of the dispersion correction.

\begin{figure*}[t!]
\includegraphics[height=\textwidth,angle=90]{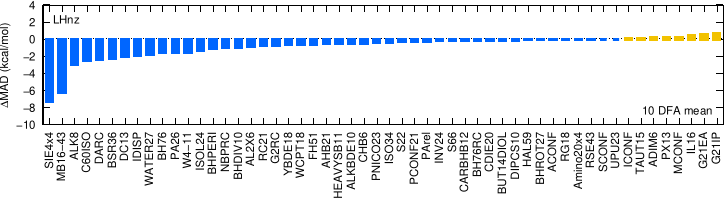}

\includegraphics[height=\textwidth,angle=90]{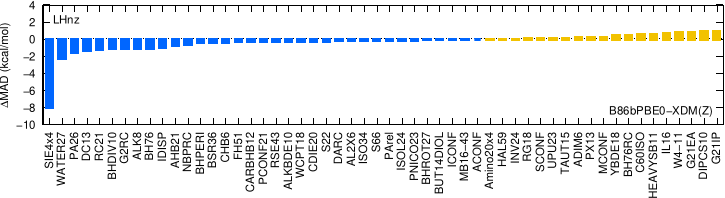}

\includegraphics[height=\textwidth,angle=90]{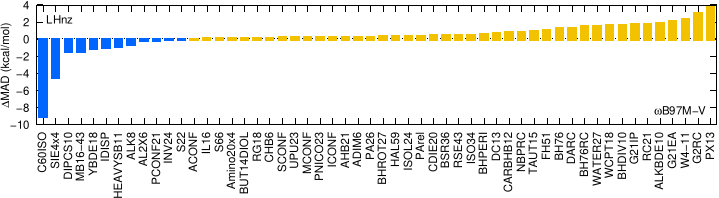}

\includegraphics[height=\textwidth,angle=90]{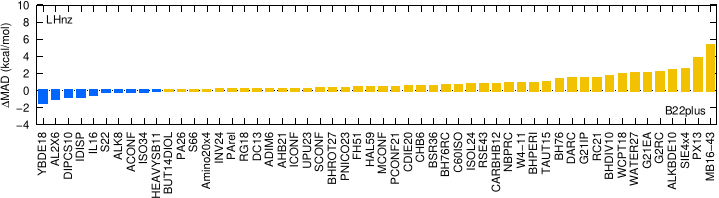}

\caption{Performance of LHnz relative to the 10-DFA mean, as well as the B86bPBE0-XDM(Z),  $\omega$B97M-V, and B22plus functionals. Blue bars indicate benchmarks for which LHnz gives a lower MAD.}\label{f:bargraphs}
\end{figure*}

Figure~\ref{f:bargraphs} shows a visualisation of the MAD results that highlights the difference in performance of LHnz versus the 10-DFA average, the B86bPBE0-XDM(Z) global hybrid, the $\omega$B97M-V range-separated hybrid functional, and the B22plus local hybrid. 
Comparison to the 10-DFA average reveals consistently good performance for LHnz for the majority of the benchmarks. Notably, LHnz gives particularly low errors for the self-interaction error (SIE4x4) and mindless (MB16-43) benchmarks, although it underperforms for ionic systems, including ionisation potentials (G21IP), electron affinities (G21EA), and ionic liquids (IL16). However, a more fair comparison is to B86bPBE0-XDM(Z), which itself already performs well relative to the 10 DFA consensus average. This comparison again shows good performance of LHnz for SIE4x4, as well also other benchmarks typically affected by delocalisation error (WATER27, PA26, DC13, RC21, BHDIV10), and confirms underperformance for G21IP, G21EA, and double ionisation potentials (DIPCS10).

Comparing to the best available RS hybrid, $\omega$B97M-V is somewhat more accurate than LHnz for the majority of the benchmarks, especially the G21EA electron affinities, W4-11 atomisation energies, G2RC reaction energies, and PX13 proton-exchange barriers. However, like most RS functionals, $\omega$B97M-V struggles for the C60ISO isomerisation energy set, and LHnz gives much better statistics for both that benchmark and for SIE4x4. Lastly, comparison to the B22plus local hybrid illustrates the limits of our existing functional form. Like many local hybrids, B22plus includes explicit non-dynamical correlation terms, which are absent from LHnz. Such correlation terms appear to greatly improve performance for the challenging MB16-43 benchmark of artificial molecules beyond the already good results seen with B86bPBE0-XDM(Z) and LHnz. The additional correlation terms also appear to be quite beneficial for the PX13, G21EA, and G21IP benchmarks, which were problematic for LHnz. Indeed, LHnz's use of full long-range exact exchange without compensating non-local correlation terms would be expected to lead to a systematic error for ionic systems\cite{kaupp2026local} that we hope to address in future work.

\begin{table*}[t!]
\caption{The top 32 global, range-separated, and local hybrid functionals sorted by WTMAD-4. 
$N_\text{param}$ indicates the number of empirical fit parameters in each functional, while NN indicates an exact-exchange mixing fraction determined from a neural network. Also shown are the numbers of outliers and the GKMTKN55 subset responsible for the maximum outlier in terms of both percent and absolute errors.  WTMAD-2 values also listed for comparison; as discussed in Ref.~\citenum{wtmad4}, the WTMAD-2 literature values may have a margin of error of $\sim$1-2\% due to ambiguity of specific reference values used and unavailability of complete benchmark results from literature. If the same functional-DC combination was listed in multiple sources, only the lowest WTMAD-4 value and corresponding outlier statistics are quoted here.  Methods listed in bold are from the present work. The B22 family of functionals implicitly include the XDM(BJ) dispersion correction, while LHz and LHnz implicitly include XDM(Z).}\label{tab:rankings}
\centering 
{\footnotesize
\setlength{\tabcolsep}{2pt}
\begin{tabular}{llr|cc|rrrl|rrrl}\hline
& & & \multicolumn{2}{c|}{WTMAD-$N$} & \multicolumn{4}{c|}{Ratio outliers} & \multicolumn{4}{c}{Difference outliers} \\
Functional & Type & $N_\text{param}$  & 2 & 4 & $N_{r>1}$ & $N_{r>2}$ & MAX $r$ & Set & $N_{d>2}$ & $N_{d>5}$ & MAX $d$ & Set       \\ 
\hline
$\omega$LH25tdE-D4\cite{kaupp-beyondzero} & RS+local & 22 &  2.57 &  3.32 & 3           &   1       &  2.55   & IL16      &  1        &  0        &  3.69   & DIPCS10   \\ 
LH24n-D4\cite{kaupp2025data,kaupp-beyondzero} & local & NN & 2.91 &  3.54 & 4           &   1       &  2.17   & IL16      &  1        &  0        &  2.27   & DIPCS10   \\
B22plus\cite{b22plus} &  local & 20 & 2.76 &  3.63 & 6           &   1       &  2.64   & IL16      &  0        &  0        &  0.91   & IL16      \\
$\omega$B97M-V\cite{wb97mv,najibi2018nonlocal} & RS & 12 & 3.08 & 3.74 & 6 & 1 & 2.22 & C60ISO    &  1 & 1 &  6.51 & C60ISO    \\
LH25nP-D4\cite{kaupp2026local} & local & NN & 2.44 & 3.79 & 6 & 3 &  3.03 & DIPCS10   & 2 & 2 &  9.67 & C60ISO    \\  
$\omega$LH23tdE-D4\cite{kauppsc,kaupp-beyondzero} & RS+local & 6 &  3.58 &  4.07 & 8            &   1       &  3.12   & IL16      &  0        &  0        &  1.18   & DIPCS10   \\
B22\cite{b22plus} & local & 9 & 3.31 &  4.21 & 6           &   0       &  1.91   & ADIM6     &  0        &  0        &  1.41   & DIPCS10   \\
$\omega$B97M-D4\cite{wb97mv,najibi2020dft} & RS & 15 & 3.69 & 4.30 & 11 & 1 & 2.24 & C60ISO    &  2 & 2 &  6.60 & C60ISO    \\
$\omega$B97X-V\cite{wb97xv,najibi2018nonlocal} & RS & 10 & 3.84 & 4.44 & 5 & 1 & 2.57 & C60ISO    &  2 & 2 & 13.21 & MB16-43   \\
$\omega$B97M-D3(BJ)\cite{wb97mv,najibi2018nonlocal} & RS & 14 & 3.77 & 4.61 & 11 & 3 & 2.47 & C60ISO    &  2 & 1 &  7.86 & C60ISO    \\
LH20t-D4\cite{lhleftright,kaupp-earlylh} & local & 12 &  4.46 &  4.62 & 5           &   1       &  2.29   & IL16      &  1        &  0        &  2.79   & DIPCS10   \\
B22-10b\cite{b22} & local & 7 &  3.60 &  4.68 & 8           &   1       &  3.12   & ADIM6     &  0        &  0        &  1.88   & HEAVYSB11 \\
$\omega$B97X-D4\cite{wb97xv,najibi2020dft} & RS & 13 & 3.96 & 4.63 & 8 & 1 & 2.65 & C60ISO    &  1 & 1 &  8.83 & C60ISO    \\
$\omega$B97M-D3(0)\cite{wb97mv,najibi2018nonlocal} & RS & 14 & 4.10 & 4.73 & 9 & 1 & 2.31 & C60ISO    &  2 & 1 &  7.01 & C60ISO    \\
scLH22t-D4\cite{kaupp-earlylh,kaupp-earlylh} & local & 14 &  4.37 &  4.71 & 7           &   1       &  2.58   & IL16      &  2        &  0        &  3.07   & DIPCS10   \\
B22-10a\cite{b22} & local & 7 &  3.70 &  4.83 & 11           &   2       &  2.31   & ADIM6     &  2        &  0        &  2.55   & DIPCS10   \\
LH24x-D4\cite{kaupp-nogauge} & local & 11 & 4.83 &  5.09 & 10           &   3       &  3.07   & IL16      &  0        &  0        &  1.16   & DIPCS10   \\
LH24x-core-D4\cite{kaupp-nogauge} & local & 9 & 4.91 &  5.18 & 10           &   3       &  3.24   & ADIM6     &  0        &  0        &  1.45   & DIPCS10   \\
\textbf{LHnz} & local & 3 & 4.90 &  5.19 & 9           &   1       &  2.03   & ADIM6     &  0        &  0        &  0.73   & G21IP     \\
PW6B95-D3(BJ)\cite{pw6b95,najibi2018nonlocal} & global & 8 & 5.45 & 5.19 & 14 & 1 & 2.31 & IL16      &  0 & 0 &  1.76 & SIE4x4    \\
M05-2X-D3(0)\cite{m052x,wtmad4} & global & 25 & 4.44 & 5.21 & 14 & 1 & 2.08 & ADIM6     &  3 & 1 &  7.02 & MB16-43   \\
M06-2X-D3(0)\cite{m062x,wtmad4} & global & 35 & 4.72 & 5.22 & 9 & 2 & 3.50 & HEAVYSB11 &  1 & 1 &  5.83 & HEAVYSB11 \\
LH24x-SC-D4\cite{kaupp-nogauge} & local & 9 &  5.05 &  5.22 & 9           &   3       &  3.76   & ADIM6     &  0        &  0        &  1.82   & SIE4x4    \\
$\omega$B97X-D3(BJ)\cite{wb97xv,najibi2018nonlocal} & RS & 12 & 4.07 & 5.26 & 9 & 2 & 2.99 & C60ISO    &  3 & 1 & 10.65 & C60ISO    \\
PW6B95-V\cite{pw6b95,najibi2018nonlocal} & global & 6 & 5.50 & 5.28 & 10 & 2 & 2.49 & ADIM6     &  0 & 0 &  1.81 & SIE4x4    \\
\textbf{LHz} & local & 2 &  5.10 &  5.30 & 10           &   1       &  2.03   & ADIM6     &  0        &  0        &  0.64   & G21IP     \\ 
PW6B95-D3(0)\cite{pw6b95,najibi2018nonlocal} & global & 9 & 5.50 & 5.33 & 13 & 1 & 3.18 & ADIM6     &  0 & 0 &  1.83 & SIE4x4    \\
revPBE0-D3(BJ)\cite{zhang1993comment,perdew1996generalized,adamo1999toward,xdmz} & global & 2 & 5.21 & 5.34 & 12 & 0 & 1.51 & W4-11     &  1 & 0 &  3.03 & W4-11     \\ 
GH-PBEx-B95c-D4\cite{kaupp-nogauge} & global & 8 &  5.24 &  5.35 & 13           &   1       &  2.20   & ADIM6     &  0        &  0        &  1.11   & DIPCS10   \\
revPBE0-XDM(Z)\cite{zhang1993comment,perdew1996generalized,adamo1999toward,xdmz} & global & 1 & 5.29 & 5.38 & 14 & 0 & 1.48 & W4-11     &  1 & 0 &  2.83 & W4-11     \\ 
$\omega$B97X-D3(0)\cite{wb97xd3,najibi2018nonlocal} & RS & 12 & 4.65 & 5.43 & 15 & 1 & 2.54 & C60ISO    &  2 & 2 & 17.20 & MB16-43   \\
\textbf{B86bPBE0-XDM(Z)}\cite{becke1986large,perdew1996generalized,price2023xdm,xdmz} & global & 1 & 5.55 & 5.44 & 16 & 0 & 1.18 & RSE43 & 0  & 0 & 0.61  & SIE4x4  \\
\hline
\end{tabular}\\
}
\end{table*}

Finally, we compare the overall performance of LHz and LHnz to the best available  hybrids in the literature. Table~\ref{tab:rankings} ranks the top 32 (out of 114) dispersion-corrected global, range-separated, and local hybrid functionals by WTMAD-4, although the corresponding WTMAD-2 is also shown for comparison. Note that hybrid functionals that use neural-network architecture beyond the mixing function, such as DM21,\cite{kirkpatrick2021pushing} are omitted from the present analysis.

This is the first time that WTMAD-4 has been evaluated for the given local hybrids, and the results re-emphasise the issues seen with WTMAD-2 overweighting specific benchmarks.\cite{wtmad4} 
For example, one can compare the results for the two functionals with neural-network mixing functions, LH24n-D4 and LH25nP-D4. The latter very heavily weighted the BH76 reaction barrier set in its training and also included the WTMAD-2 of a reduced ``diet-GMTKN55'' set in its loss function.\cite{kaupp2026local} This lead to a considerably lower WTMAD-2 value than seen with LH24n-D4, although the latter gives a lower WTMAD-4. Similarly, LH20t-D4 and $\omega$B97M-D3(BJ) give essentially identical WTMAD-4 values, despite the latter functional having a much lower WTMAD-2.

The outlier analysis in Table~\ref{tab:rankings} informs which benchmarks are the most challenging for the various functional classes. As noted previously,\cite{wtmad4} the C60ISO benchmark is problematic for the range-separated hybrids, likely due to the dependence of the optimal $\omega$ parameter on system size.\cite{bredas,isborn,whittleton} Conversely, the local hybrids tend to perform well for C60ISO, but have the double ionisation potentials (DIPCS10) as a consistent outlier.\cite{kaupp2026local} Intermolecular binding energies of ionic liquids (IL16) and alkane dimers (ADIM6) also frequently appear as ratio outliers; however, this does not mean that the local and global hybrid functionals have large errors for these sets so much as that the 10 reference DFAs already perform exceptionally well. For example, the 10-DFA consensus average gives a MAD of 0.17 kcal/mol for ADIM6, which is particularly low, and doubling this value would still be an acceptable error for non-covalent interactions. Similarly, for the IL16, the 10-DFA consensus average gives a MAD of 0.55 kcal/mol relative to a $|\overline{\Delta E}|$ of 109 kcal/mol; again, doubling this error would be quite acceptable for ionic bonding. 

Comparing LHnz with the best available hybrids, Table~\ref{tab:rankings} shows that it gives comparable performance to the 
LH24x-D4 and LH24x-core-D4 methods in terms of WTMAD, but with lower maximum outliers. Like LHnz, both LH24x-D4 and LH24x-core-D4 make use of the density-functional exchange-energy density in their mixing fraction definition and do not require a gauge calibration function.\cite{kaupp-nogauge} 
Impressively, LHnz gives a lower \mbox{WTMAD-4} than any global hybrid functional and outperforms functionals with 2--11$\times$ as many parameters. While far lower WTMAD values can be achieved with the best local and range-separated hybrids, these tend to be fit extensively to thermochemical benchmarks. In terms of functionals that involve both few parameters and low outliers, $\omega$LH23tdE-D4 ($N_\text{param}=6$), LHnz ($N_\text{param}=3$), and B86bPBE0-XDM(Z) ($N_\text{param}=1$) are among the top performers. This finding indicates the importance of dispersionless exchange, and that inclusion of sound physics is able to offset accuracy gains from parameterisation. Although LHnz does not approach the accuracy of the very best hybrid functionals, it is notable for having fewer parameters than any functional with a lower WTMAD-4; the same may also be said for LHz and revPBE0-XDM(Z).

\section{Summary}

Many density functional approximations have been developed using a data-driven framework, with fits to increasingly large sets of molecular reference data. However, we have recently advocated the benefits of minimally empirical functionals based on sound physics, such as the use of dispersionless exchange functionals to avoid error cancellation with the dispersion correction.\cite{requirements,xdmz} In the present work, we extend this design philosophy to local hybrid functionals based on the B86bPBE0-XDM(Z) global hybrid. Two new local hybrid functionals were proposed, termed LHz and LHnz, which involve only two and three empirical parameters, respectively. LHz defines the local mixing fraction in terms of the correlation length via an error function. LHnz is similar, but introduces a multiplicative scaling factor that depends on the effective exchange--correlation hole normalisation to the error function argument. Both are shown to yield significantly improved performance on the GMTKN55 molecular benchmark than the parent B86bPBE0-XDM(Z) functional, while LHnz is more accurate than LHz for open-shell systems with inherent delocalisation error.

Comparison with literature functionals shows that LHnz gives a lower WTMAD-4 than all global hybrids, with no large outliers. While eighteen local and range-separated hybrid functionals yield lower WTMAD-4 values, all rely on 2--11$\times$ as many parameters (or neural networks). LHnz has the further benefit of conceptual simplicity, as it does not involve range separation, calibration functions, or power-series expansions. However, similar to other local hybrids, LHnz tends to underperform for benchmarks involving ionic systems. Improved correlation models will be required to improve accuracy further and we plan to explore this in future work.

\section*{Acknowledgements}

KRB and ERJ thank the Natural Sciences and Engineering Research Council (NSERC) of Canada for financial support and the Atlantic Computing Excellence Network (ACENET) for computational resources. ERJ additionally thanks the Royal Society for a Wolfson Visiting Fellowship, and Dr.\ S.\ G.\ Dale for assistance with the postG program.

\section*{Data Availability Statement}

The data that support the findings of this study are available in the supplementary information.

\section*{Conflicts of Interest}

There are no conflicts to report.


\bibliography{refs.bib}

\end{document}